\documentclass[10pt, twoside]{article}
\usepackage[utf8]{inputenc}
\usepackage[T1]{fontenc}
\usepackage[hypertexnames=false]{hyperref}
\usepackage{bm}
\usepackage{subcaption}
\usepackage{siunitx}
\usepackage[pdftex]{graphicx}
\graphicspath{ {./figs/} }
\usepackage{amsmath}
\usepackage{mathrsfs}
\usepackage{amsfonts}
\usepackage{amssymb}
\usepackage{mathtools}
\usepackage{array}
\usepackage{fancybox}
\usepackage{bm}
\usepackage{upgreek}
\usepackage{xcolor}
\usepackage[table]{xcolor}
\usepackage{todonotes}
\usepackage{lineno}
\usepackage{comment}
\usepackage{makecell}
\usepackage{cancel}
\usepackage{empheq}
\usepackage{setspace} 
\usepackage{nicefrac}
\usepackage{booktabs}

\usepackage{makecell}
\usepackage[letterpaper,twoside,textwidth=18cm,textheight=25cm]{geometry}
\usepackage{biblatex}
\newcommand{\teff}{\ensuremath{t_{\mathrm{eff}}}}
\newcommand{\eeff}{\ensuremath{e_{\mathrm{eff}}}}

\newcommand{\footremember}[2]{%
	\footnote{#2}
	\newcounter{#1}
	\setcounter{#1}{\value{footnote}}%
}
\newcommand{\footrecall}[1]{%
	\footnotemark[\value{#1}]%
}

\title{Comparison of numerical schemes for integration of tractions in discrete models for fracture}

\author{Monika Středulová\footremember{Brno}{Brno University of Technology, Faculty of Civil Engineering, Brno, Czechia}\footremember{email}{Corresponding author: monika.stredulova@vut.cz} and Gianluca Cusatis\footremember{Northwestern}{Northwestern University, Department of Civil and Environmental Engineering, Evanston, IL USA} and Jan Eliáš\footrecall{Brno}}
\date{}

\begin{document}
\maketitle

\section*{Abstract}

This numerical  study investigates various types of numerical integration techniques in mechanical lattice-type discrete models for concrete. Using the Lattice Discrete Particle Model (LDPM) as a~reference, the geometry is reformulated into a~set lattice elements connecting two particles with an~associated area that represents mutual particle contact. The original LDPM constitutive functions are preserved, including the effect of volumetric strain, while different integration schemes, ranging from single- to multi-point formulations, are introduced to control bending stiffness emerging from linear profile of normal contact traction. The integration schemes are assessed through simulations of a~dogbone tension test, three-point bending, unconfined compression, and hardening response. Results show that tensile, bending, and hardening behavior are largely insensitive to the integration choice. In contrast, the unconfined compression is strongly affected: single-point integration cannot account for non-constant traction profile and therefore underestimates model resistance, whereas multi-point integration yields consistent predictions. Simulations with artificially constrained rotations further confirm the decisive role of correct integration of non-constant traction profile. The findings underline that an~adequate representation of rotations and aggregate interactions is essential for reliable modeling of compression-dominated behavior in lattice-type frameworks.

\section*{Keywords}
Lattice Discrete Particle Model; concrete; fracture; volumetric strain; integration scheme; contact stress


\section{Introduction}


Concrete and other quasi-brittle materials exhibit complex failure mechanisms governed by heterogeneity across multiple length scales. Capturing these mechanisms in numerical simulations is essential to predict structural behavior under different loading conditions \parencite{Hre-41}. Homogeneous continuum models, while efficient and scalable, often struggle to reproduce the important effect of material heterogeneity on crack nucleation, strain localization, fracture, and damage. This has motivated the development of mesoscale  discrete models with \emph{physical} discretization, which (i) explicitly represent the underlying material heterogeneity and (ii) naturally allows initiation and growth of discrete oriented cracks~\parencite{BolEli-21}.

A broad family of such models has been proposed in the literature, including truss lattices, beam lattices, and rigid-body spring models~\parencite{NitTej24,GanYan-24,GuLuk26,ZhaFas-26,QiuLuo-26,KarNak-26}. These approaches share the ability to capture fracture and heterogeneity, yet differ substantially in how particle interactions, rotational degrees of freedom, and integration schemes are handled.

Within this family, the Lattice Discrete Particle Model (LDPM) has emerged as a~particularly successful framework for simulating the mechanical response of concrete at the mesoscale. By representing aggregates as rigid particles connected through interacting facets, LDPM can reproduce tensile cracking, shear friction, and compressive pore collapse. The LDPM compressive formulation based on the dependence to \emph{volumetric strain} is unique among discrete models and it is of quintessential importance to capture frictional effects under confinements. The volumetric strain cannot be estimated from single two-node interaction; therefore LDPM, contrarily to most other discrete models, is formulated within a~system of tetrahedral elements each connecting four particles and the volumetric strain is related to the relative change of the tetrahedron volume.  LDPM has been validated in a~wide range of loading scenarios and has been extended to, e.g, multiscale~\parencite{YanYin-26} and multiphysics formulations~\parencite{SheZha-27} and used to simulated real structural components~\parencite{AlnPel-19,BhaGom-21,TroLal-25}. One downside if this approach is that for each LDPM tetrahedron the constitutive equations must be evaluated multiple time, namely 12 times for the specific geometrical representation of the material internal structure adopted in LDPM \parencite{CusPel-11, CusMen-11}.


The most recent effort to reduce the computational cost of LDPM has been presented by Wang and co-workers in two consecutive papers. In Ref.~\parencite{Wang2024}, the authors introduce simplified LDPM geometries: (i) an~edge-based formulation in which multiple facets are regrouped into a~single projected area around the edge between two particles (essentially identical to the existing confinement–shear lattice model \parencite{CusBaz-06}), and (ii) a~reduced six-facet LDPM configuration. Both are compared against the standard twelve-facet variant \parencite{CusPel-11, CusMen-11}. In their follow-up study~\parencite{Wang2025}, an~adaptive scheme is proposed in which simulations begin with the edge-based geometry and transition to the six- or twelve-facet formulation once a~user-defined stress-limit ratio is exceeded. This approach provides substantial computational savings.

Both publications offer valuable observations. For most loading scenarios studied in Ref.~\parencite{Wang2024} (biaxial and triaxial compression, torsion, cyclic compression, and three-point bending) the simplified geometries match the twelve-facet LDPM response closely. The only exception is unconfined compression, where the edge-based formulation performs poorly. The authors attribute this to the reduced number of facets in the edge-based formulation.

However, neither publication expands on this explanation nor investigates the interaction mechanics responsible for the observed behavior. Given that the six-facet and twelve-facet models are reported as ``indistinguishable'', facet count alone cannot explain the deterioration of the ability to simulate unconfined compression tests.

 \emph{Central hypothesis:} In the present article, we propose and numerically examine the following hypothesis. When a~single integration point is used to integrate normal tractions at the interparticle contact, the couple arising from the non-constant traction profile is lost. By adding multiple integration points, normal tractions acting at these points are capable to account for these couples and, consequently, develop some curvature resistance. This emerging bending stiffness is omittable in tensile and bending loading scenarios, but critical in simulations of unconfined compression. This interpretation aligns with the empirical observations of Refs.~\parencite{Wang2024, Wang2025}, although the mechanism has not been explicitly analyzed in the prior literature.

This hypothesis is examined by establishing a~theoretical and numerical framework linking the numerical integration scheme to the bending behavior of lattice-model particle interactions. We examine how the number and distribution of integration points control the emergence of bending resistance in an~edge-based formulation and quantify the resulting influence on tensile, flexural, and, most notably, compression-dominated loading paths. This provides a~systematic interpretation of when simplified interaction geometries remain adequate and when they fundamentally alter the underlying mechanics.The numerical studies are performed on a~decomposed four-node LDPM tetrahedrons modified to a~two-node lattice contacts, including a~technique to evaluate \emph{volumetric} strain needed in confined compression simulations. Furthermore, we present a~comprehensive analysis of the computational performance of different traction integration variants. 

\section{Modeling approach}

The following subsections summarize the reference LDPM formulation in terms of geometry and constitutive laws. These definitions provide the baseline for the geometric decomposition and integration variants investigated in Sections 2.2 and 2.3.

\subsection{Brief review of the Lattice Discrete Particle Model}\label{212}

After defining the geometric domain of a~specimen or a~structure to be modeled, spherical particles are generated and randomly placed within the domain. The particle size distribution is prescribed by a~cumulative distribution function that reflects the actual aggregate grading~\parencite{YanTro-24}, thereby providing a~physically representative description of the concrete mesostructure.

Then a~tetrahedralization is performed using particle centers, producing the basic element of the model: a~tetrahedron. The vertices of the tetrahedrons are hereafter referred to as nodes. Each node possesses six degrees of freedom (DoF), three translational, and three rotational.

Each tetrahedron represents the interaction of four particles (aggregates) through twelve internal triangular facets that transfer forces via the embedding matrix. A~single facet corresponds to the contact interface between two particles; there are two such facets between each pair of particles within one tetrahedron(see triangles a \& b in Fig.~\ref{geom_creation}a). Strains and tractions are evaluated at the facet level at \emph{integration points} (IP). Their governing equations are described in the following subsection. Figure~\ref{geom_creation}a illustrates a~tetrahedron with its12 facets. In the following, this geometrical representation is referred to as \texttt{tetra}.

In LDPM, the constitutive behavior is defined at the facet level. Strains are obtained from the displacement jump between the two nodes adjacent to a~facet, projected into the local coordinate system formed by the facet normal and two orthogonal tangential directions. Consequently, three strain components are defined: the normal strain $e_N$, and the two shear strains $e_M$ and $e_L$. The corresponding tractions are denoted $t_N$, $t_M$, and $t_L$. In the elastic regime, the response is linear and defined as $t_N = E_{0} e_N$, $t_M = \alpha E_{0} e_M$ and $t_L = \alpha E_{0} e_L$. The first elastic parameter, $E_0$, is the effective normal modulus, the second parameter, $\alpha$, relates the normal and tangential elastic response on the contact. Approximate relations of $E_0$ and $\alpha$ to Young's modulus and Poisson's ratio can be found in Refs.~\parencite{CusPel-11,Eli20}.

Further, the inelastic response is based on effective traction and is asymmetric, depending on whether there is tension or compression in normal directions~\parencite{CusPel-11}. Under tension, both normal and shear tractions are dictated by the effective traction, that is incrementally elastic within the domain bounded by function $\sigma_{bt}$. This boundary function simulates softening in tension and is dependent on strain history and current straining direction. In compression, the normal response accounts for pore collapse and yielding at low confinement, and for compaction and hardening at higher confinement. The compressive boundary $\sigma_{bc}$ depends on the volumetric strain $\varepsilon_{V}$ and the deviatoric strain $\varepsilon_{D}$, with their ratio governing the transition between softening and hardening. Shear tractions under compression are governed by a~frictional law that couples the tangential response to the normal traction. The response is linearly elastic at small strains, limited by a~Mohr--Coulomb type criterion, and evolves into frictional sliding beyond the peak strength. The formulation of the LDPM constitutive equations introduced in Ref.~\parencite{CecSal-17} is adopted in this study.

In summary, the LDPM constitutive laws capture tensile fracture, shear friction, and compressive pore collapse at the mesoscale. A~detailed description of the mathematical formulation is provided in Appendix~\ref{app:constitutive}, while calibration procedures and validation in various loading scenarios can be found in the original publications  \parencite{CusMen-11, CusPel-11, CecSal-17}. In the present study, these constitutive relations are adopted without modification; the focus is placed on the geometric representation of the model and the associated integration strategies.

\subsection{Geometry decomposition and volumetric strain}

\begin{figure}[h]
    \centering
    \includegraphics[trim={0 0.5cm 0 0},clip, width=\textwidth]{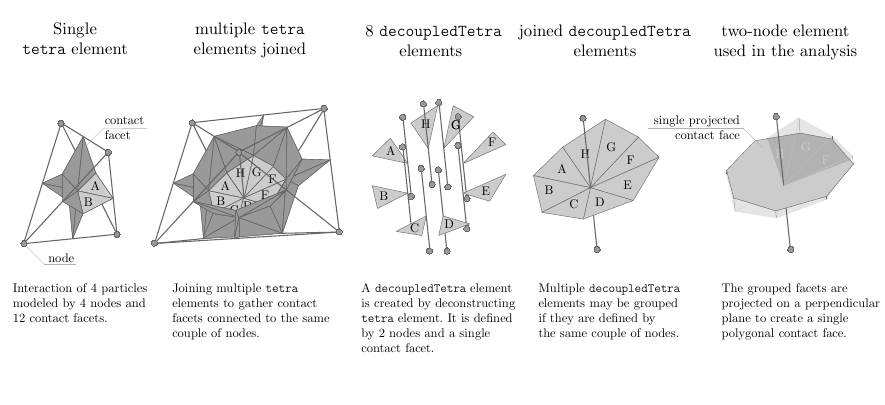}
    \caption{ Decoupling of an~LDPM \texttt{tetra} element into individual \texttt{decoupledTetra} elements, which are subsequently regrouped and projected to form two-node lattice elements.}
    \label{geom_creation}
\end{figure}

This subsection describes the transformation of the LDPM \texttt{tetra} element into an~equivalent lattice representation, in order to investigate mutual particle interaction and, in particular, the role of particle rotations. The decomposition is illustrated in Fig.~\ref{geom_creation}.  A~\texttt{tetra} element(Fig. ~\ref{geom_creation}a) models the interaction of four particles by twelve triangular facets, and can therefore be decomposed into twelve lattice elements. Each lattice element is defined by two nodes and one triangular contact facet inherited from LDPM geometry. This geometry is used as a~second reference solution and is referred to as \texttt{decoupledTetra}(Fig. ~\ref{geom_creation}c).

The \texttt{decoupledTetra} elements sharing the same pair of nodes may be regrouped(Fig. ~\ref{geom_creation}b, c and d) . As shown in Fig.~\ref{geom_creation} b, facets from different \texttt{tetra} elements that share two coinciding nodes are merged and projected onto a~plane perpendicular to the line connecting the two nodes, forming a~new two-node lattice element with a~single polygonal contact facet(Fig. ~\ref{geom_creation}e).

The original position of the facets is kept, and the re-grouping is performed to allow the facets to exhibit different mechanical behavior, while preserving the model topology. The number of connected facets depends on (i) the number of neighboring \texttt{tetra} elements and (ii) whether the element lies on the boundary of the domain. Since neither the number nor the placement of nodes changes during this regrouping, the number of degrees of freedom remains identical to that of the original LDPM geometry.

As previously mentioned, critical part of concrete behavior in compression is the dependency on the \emph{volumetric strain}. The constitutive functions of LDPM incorporate this effect and the volumetric strain is evaluated from the change in tetrahedral volume. After decomposing the \texttt{tetra} elements into two-node contacts, the volumetric strain should also be evaluated. For this reason, the tetrahedralization of the domain is kept (or newly created if the lattice model is not transformed from LDPM \texttt{tetra} objects) and the volumetric strain for each auxiliary tetrahedron is evaluated from the translations of the four nodes forming it  
. Each two-node lattice element is attached to several auxiliary tetrahedrons and volumetric strain for all integration points inside the element is obtained as their average volumetric strain. Different strategies can be implemented, for example, weighted average based on tetrahedral volumes.
However, it is our expectation they have only limited influence on the results. The strategies to evaluate the volumetric strain are therefore not studied and compared here: we adopt a~simple average from the tetrahedrons attached to the a~given two-node lattice element.


\subsection{Integration of tractions over the contact area}

For the two reference geometries (\texttt{tetra} and \texttt{decoupledTetra}), the same integration rule is employed: a~single integration point is placed at the centroid of each triangular facet. This approach has been used in LDPM from its beginning. Since the interaction between two aggregates is represented by multiple triangular facets (see the first two sketches of \texttt{tetra} elements in Fig.~\ref{geom_creation}), a~certain bending resistance emerges, as each facet can develop an~independent normal traction to resist relative particle rotation.

 To integrate traction in the two-node lattice elements, the contact face (i.e., the projected area) is subdivided into triangular facets by introducing some central point. Each triangle is defined by this central point and two vertices on the boundary of the contact face. Two strategies are considered: (i) the central point is placed at the \emph{centroid} of the contact face (denoted by \texttt{C}), and (ii) the central point corresponds to the original one of the decomposed \texttt{tetra} elements and lies on the \emph{intersection}of the the contact plane with the line connecting the nodes (denoted by \texttt{I}).

Based on this subdivision, the same integration rule as for the \texttt{tetra} and \texttt{decoupledTetra} elements is applied by placing a~single integration point at the centroid of each triangle. This integration scheme is hereinafter denoted as \texttt{\_1T}. In addition, two alternative integration rules are introduced for the lattice element: (i) a~single integration point located at the central point of the polygonal face (\texttt{\_1}), and (ii) three integration points distributed within each triangle (\texttt{\_3T}).

\begin{figure}[ht!]
    \centering
    \includegraphics[trim={0 0 0 0},clip, width=\textwidth]{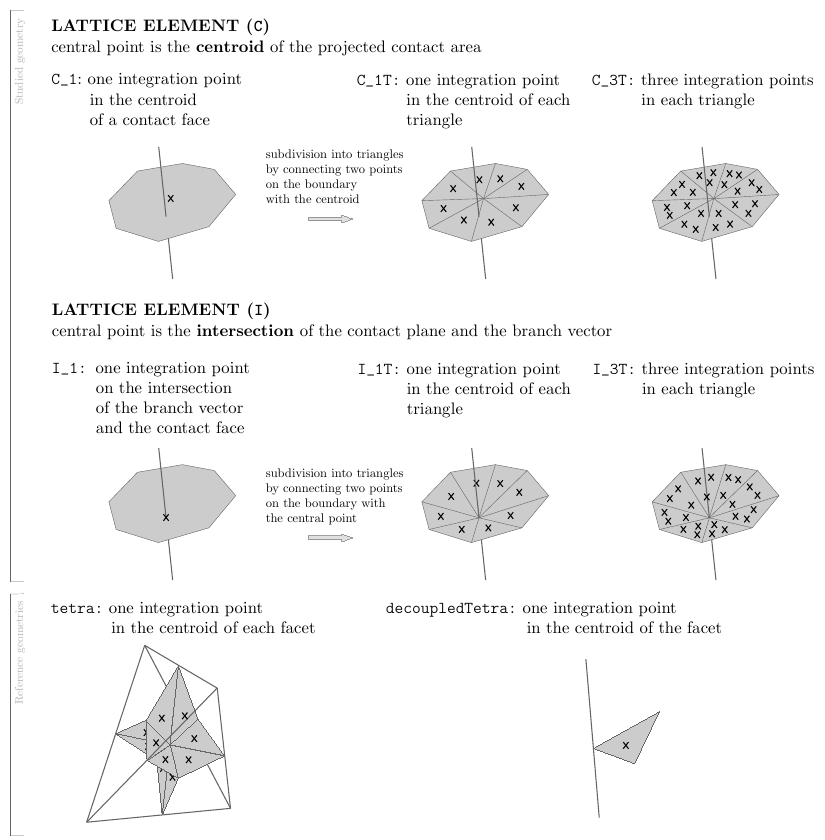}
    \caption{Geometries and integration rules used in the numerical analysis.}
    \label{IPs}
\end{figure}

By combining the two possible locations of the central point with the three integration rules, a~total of six variants are obtained. Three integration schemes are based on the centroidal position: \texttt{C\_1}, \texttt{C\_1T}, and \texttt{C\_3T}, while the remaining three schemes use the intersection-based original central point: \texttt{I\_1}, \texttt{I\_1T}, and \texttt{I\_3T}. The integration techniques are summarized in Tab.~\ref{table_IPs} and illustrated in Fig.~\ref{IPs}.

Placing the central point at the centroid is motivated by classical lattice-particle models based on Voronoi or power tessellations, where tractions are typically integrated using a~single integration point located at the centroid of the contact face~\parencite{ParCho-24,RaiZha-25,QiuLuo-25}. Moreover, the intersection of the line connecting nodes 
with the contact plane does not necessarily lie within the contact area in Voronoi-based models. The centroids are expected to provide a~more representative information about the polygonal face.

\begin{table}[h!]
\centering
\begin{tabular}{lccc}
\toprule
\textbf{variant} & \textbf{central point} & \textbf{number of IPs}\\
\midrule
\texttt{C\_1}   & centroid        & 1 per polygon          \\
\texttt{C\_1T}  & centroid        & 1 per triangle         \\
\texttt{C\_3T}  & centroid        & 3 per triangle         \\
\texttt{I\_1}   & intersection & 1 per polygon        \\
\texttt{I\_1T}  & intersection & 1 per triangle       \\
\texttt{I\_3T}  & intersection & 3 per triangle       \\[0.1cm]
\texttt{tetra}  &  & 1 per triangle       \\
\texttt{decoupledTetra}  &  & 1 per triangle      \\\bottomrule
\end{tabular}
\caption{Summary of integration rules applied on facet polygons.}
\label{table_IPs}
\end{table}

The mechanical implications of these integration variants are directly related to bending resistance at the contacts. Using a~single integration point per polygon (\texttt{C\_1} or \texttt{I\_1})effectively reduces linear profile of normal tractions to the  uniform one, which provides no (\texttt{I\_1}) or only minimal (\texttt{C\_1}) resistance to curvature, and therefore typically results in a~softer response. Such a~formulation can adequately capture normal and shear interactions between particles, but omits the bending effect. The centroidal formulation (\texttt{C\_1}) has some eccentricity of the central IP and therefore provides partial resistance to rotational difference between particles; however the component arising from non-uniform tractions is still missing. Increasing the number of integration points enhances rotational resistance by improving the description of normal traction over the contact area, while preserving the model’s ability to transmit normal and shear loading. Consequently, the whole discrete model becomes stiffer in elastic regime. This has already been shown by~\textcite[Fig.~5]{Eli17} in two dimensions where single IP or 5 IPs were considered.

Note that another option to bring resistance to mutual particle rotations into the model is to directly include rotational stiffness into the constitutive model. The curvature in lattice-particle models is evaluated as the discrete rotational jump between particles divided by the contact length. Then, the couples obtained by multiplying the curvature with the elastic or inelastic material constants are imposed on the particles~\parencite{AsaAoy-17,ZhaFas-25,EliCus25}. In classical lattice models the rotational stiffness is provided by Euler-Bernoulli beams connecting the particles~\parencite{SagPra-19,ChaLia-22,SayCha-23}.
However, the direct introduction of bending stiffness complicates the formulation of the constitutive equations in the nonlinear regime,because one needs to develop coupling between normal, shear and bending components. Generating the bending moment via multiple IPs does not affect the constitutive functions, although the magnitude of the moment cannot be modified independently.

\section{Numerical analysis}

A series of simulations are run to compare the effect of integration rules and thus the importance of local facet curvature resistance to the overall specimen response. Experiments were chosen to cover different loading scenarios: tensile loading (dogbone tension test, three-point bending test) and compressive loading (unconfined compression with different boundary conditions).

The simulation setups were taken from the publication~\parencite{LalEli-25}.  All simulations used the same material and particle generation parameters as in Ref.~\parencite{LalEli-25}. Material parameters were taken from \parencite{CusMen-11} and the full table is given in the Appendix in Table \ref{table_params}. Fuller curve exponent of 0.4 was used for particle generation along with maximum aggregate size of 12 mm, cement content of 375 kg/m$^{3}$, and water-to-cement ratio of 0.5. No separate convergence study with respect to particle size was performed, as the adopted particle distribution was taken from a~previously validated model and kept identical for all studied schemes. A~finer particles would alter the mesostructural nature of the model and change its response. However, the missing bending resistance would not be restored by the finer particles results would not be qualitatively improved.

The simulations are performed using an~in-house open-source solver OAS\footnote{Open Academic Solver, \url{https://gitlab.com/kelidas/OAS}}, which was designed for mechanical and multiphysical problems. Implicit, quasi-static solver is employed. The details of each setup are described in following subsections and material parameters are given in the annex of the paper.

\subsection{Dogbone tension test}
\begin{figure}[ht!]
    \centering
    \includegraphics[trim={0 0.5cm 0 0.5cm},clip, width=\textwidth]{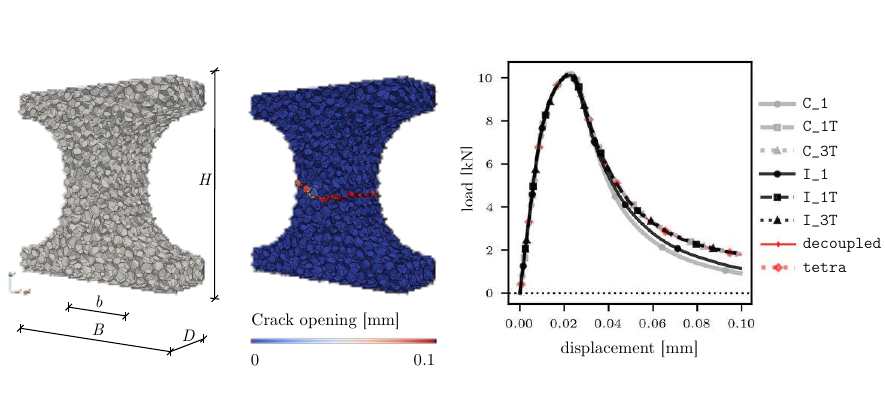}
     \caption{Dogbone tension test: geometry (left), crack pattern (middle), load-displacement curves (right).}
     \label{dogbone}
\end{figure}

A dog-bone specimen is subjected to uniaxial tension, with fracture expected to develop in the reduced cross-section. The geometry of the specimen is shown in Fig.~\ref{dogbone}, with dimensions $H = B = \qty{150}{mm}$ and $b = D = \qty{50}{mm}$.

Boundary conditions are imposed as follows. Nodes on the bottom face are fixed in all translational degrees of freedom and in all rotations. In addition, mutual translation and rotation of these boundary particles are constrained to prevent rigid-body distortion of the boundary layer, ensuring that the prescribed displacement is transferred uniformly into the specimen and that no spurious compliance is introduced. The only exception is the mutual rotation about the longitudinal axis $\theta_{z}$, which is left free. Nodes on the top face are constrained against lateral translation and all rotations, and the mutual movement between these nodes is treated in the same manner as for the bottom face.

Axial loading is applied by prescribing incremental displacement to all nodes of the top face in the direction of the longitudinal axis, ensuring a~purely uniaxial deformation state without eccentricity.

The constitutive behavior follows the LDPM formulation (see Sec.~\ref{212}). The only difference between the simulation runs lies in the number and spatial distribution of the integration points. The resulting crack pattern in the neck region obtained with the \texttt{I\_1} integration scheme, together with the load--displacement curves for all simulations, is shown in Fig.~\ref{dogbone}. Crack opening is computed as the combined contribution of the normal and shear components, $w = \sqrt{\,w_N^2 + w_M^2 + w_L^2\,}$.

\subsection{Three point bending test}

\begin{figure}[ht!]
    \centering
    \includegraphics[trim={0 4cm 0 0 },clip, width=\textwidth]{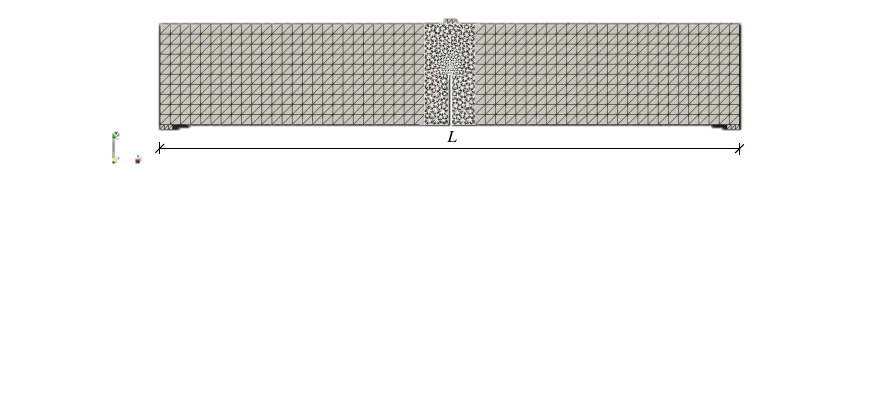}
    \caption{Three point bending test, geometry of the whole specimen, including rigid body platens.}
    \label{tpb1}
\end{figure}

The three-point bending test is simulated on the notched beam shown in Fig.~\ref{tpb1}. 
The specimen dimensions are $L = \qty{1151}{mm}$, depth $D = \qty{200}{mm}$, and width $B = \qty{100}{mm}$. 
To reduce computational cost, only the central region containing the notch is modeled with the discrete formulation, while the remaining beam sections, together with the rigid-body supports and the loading platen, are represented using standard tetrahedral finite elements. 
The width of the central discrete region is $w = \qty{100}{mm}$, and the notch has depth $a = \qty{100}{mm}$ and width \qty{4}{mm}.

Coupling between the discrete and continuum domains, as well as between the beam and the rigid platens, is achieved by ``hanging'' the particle translational degrees of freedom at the interface onto the continuum elements through their shape functions.

The beam rests on two rigid-body support platens. 
Both supports are fixed in the vertical direction and in the transverse horizontal direction, while they remain free to move along the longitudinal axis of the beam. 
A third rigid-body platen is positioned above the notch at midspan. 
This loading platen is fixed in both horizontal directions, and loading is applied incrementally via prescribed vertical displacement.

All rigid-body platens are prevented from rotating about the longitudinal axis of the beam.

\begin{figure}[ht!]
    \centering
    \includegraphics[trim={0 0 0 0 },clip, width=\textwidth]{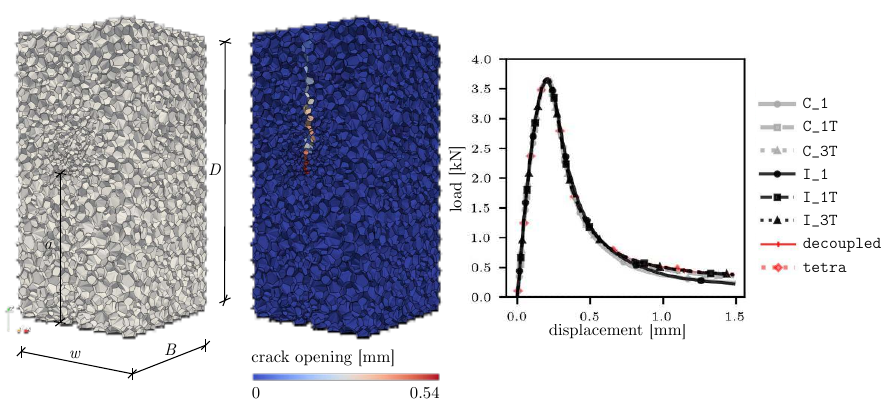}
     \caption{Three point bending test, geometry of the midlle discrete part (left), crack opening (middle) and load-displacement curves (right).}
     \label{tpb2}
\end{figure}

The resulting crack pattern obtained using the \texttt{I\_1} integration scheme, together with the load--displacement curves for all model variants, is shown in Fig.~\ref{tpb2}.

\subsection{Unconfined compression test}

\begin{figure}[ht!]
    \centering
    \includegraphics[trim={0 0 0 0},clip, width=\textwidth]{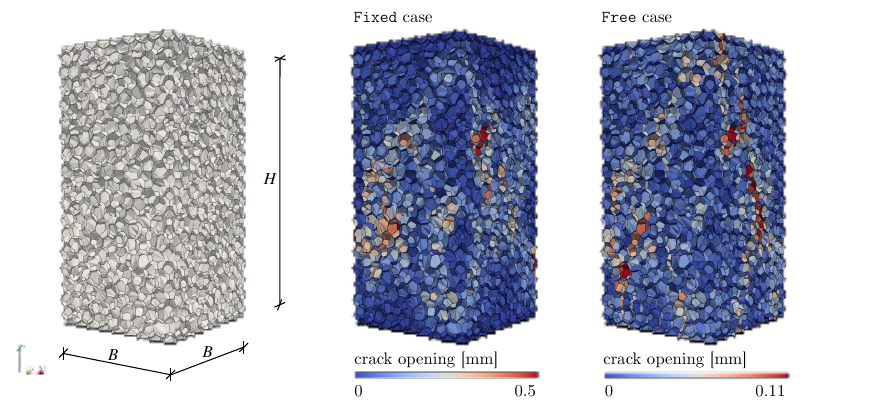}
     \caption{Unconfined compression test: geometry (left) and crack patterns of \texttt{fixed} case (middle) and \texttt{free} case (right).}
    \label{uc}
\end{figure}

A prism is analyzed under uniaxial compression to evaluate the difference in response of the discrete material with different facet integration in such scenario. The specimen dimensions are $H = \qty{200}{mm}$ and $B = \qty{100}{mm}$ in both width and depth (see Fig.~\ref{uc}, left). In real compression tests, friction develops between the loading platens and the specimen, and it can strongly influence the observed mechanical behavior \parencite{vanMier-97}. To capture the two limiting situations, the following boundary conditions are considered: (i) a~\texttt{fixed} case, representing full friction at both loading platens, where lateral movement at the specimen bases is suppressed; and (ii) a~\texttt{free} case, representing a~frictionless interface, where lateral movement of the base nodes is permitted.

In both cases, the bottom base is fixed in the vertical direction, and axial loading is introduced by prescribing incremental vertical displacement to the nodes of the top base.

In the \texttt{fixed} configuration, all nodes on the bottom base are fixed in all translational directions, and rotations about all axes are constrained. The same kinematic restrictions apply to the top base, except in the vertical direction, where displacement is prescribed to apply the load. Mutual translations and rotations between particles belonging to each base are also suppressed to ensure uniform transfer of load and to eliminate non--physical deformation modes at the boundaries.

In the \texttt{free} configuration, lateral movement is allowed for all nodes of both bases, except for one central node on each base which is fixed laterally to prevent rigid--body motion. Nodes of the bottom base remain fixed vertically, while incremental vertical displacement is prescribed to the top base. Mutual movement between particles on each base is prevented in the vertical direction and in rotations about the two horizontal axes, while in--plane horizontal translations remain unconstrained to reflect the frictionless contact condition.

\begin{figure}[ht!]
    \centering
    \includegraphics[trim={0 0.4cm 0 0.7cm},clip, width=\textwidth]{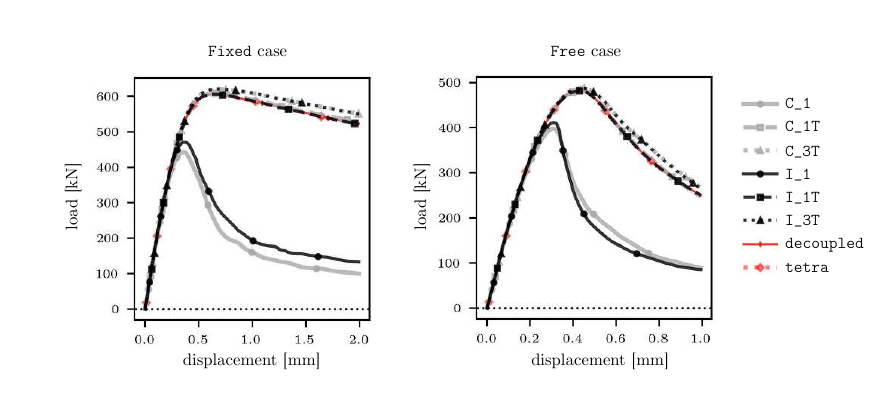}
     \caption{Resulting load-displacement curves of the \texttt{fixed} case (left) and \texttt{free} case (right).}
    \label{uc_results}
\end{figure}

Final crack patterns of both cases obtained by the \texttt{I\_1} integration are shown in the Fig.~\ref{uc}. Resulting load-displacement curves may be found in Fig.~\ref{uc_results}.

\subsection{Hardening response}
Cylinder of height $H = \qty{200}{mm}$ and base diameter $D = \qty{200}{mm}$ is subjected to confined compression to simulate the response of the material to hardening due to pore collapse.

Both bases of the cylinder are constrained against lateral translation and all rotations. The bottom base is fixed in the vertical direction, while incremental vertical displacement is prescribed to all nodes of the top base to apply the load. Mutual movement of the particles belonging to each base is suppressed in the vertical direction and in rotations about the two horizontal axes, ensuring uniform transfer of the prescribed displacement and preventing non-physical deformation of the boundary layer.

The lateral surface of the cylinder is constrained in the radial (horizontal) direction to prevent transverse expansion, and mutual lateral movement between surface particles is also prohibited to maintain a~consistent confinement condition.

\begin{figure}[ht!]
    \centering
    \includegraphics[trim={0 0 0 0},clip, width=\textwidth]{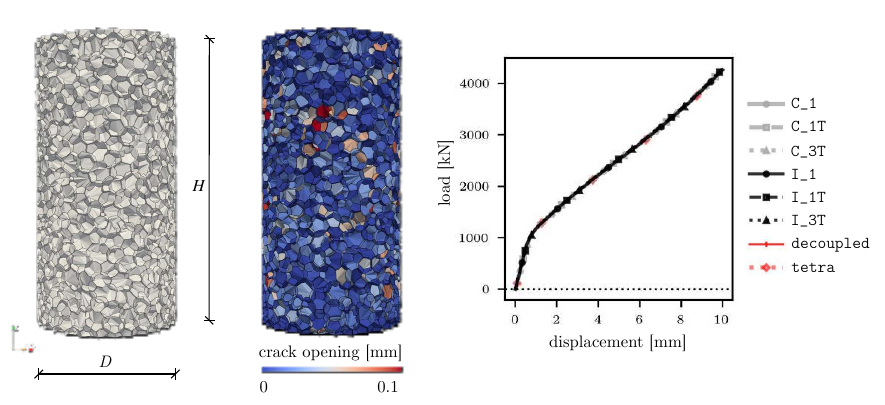}
     \caption{Response in hardening: geometry of the experiment (left), crack pattern  magnitude (middle), and load-displacement curve (right).}
     \label{us}
\end{figure}

The geometry, the final crack pattern obtained with the \texttt{I\_1} integration, and the load–displacement curves of all configurations are presented in Fig. \ref{us}. Note that the crack openings in the figure are caused mostly by shear sliding of crack faces as in the normal direction there are compressive tractions. 

\section{Discussion}
 
The relative difference in peak load, $\Delta P_{\max}$, and the normalized root mean square error of the load response, $\mathrm{NRMSE}(P)$ are shown in Tab.~\ref{table_errors}. They were evaluated as
\begin{align}
\Delta P_{\max}
&=
\frac{
\left|P_{\max}^{\mathrm{var}}\right|
-
\left|P_{\max}^{\mathrm{ref}}\right|
}{
\left|P_{\max}^{\mathrm{ref}}\right|
}
\times 100\,\% &
\qquad
\mathrm{NRMSE}(P)
&=
\sqrt{\frac{
\displaystyle\sum_{i=1}^{n}
\left(P_i^{\mathrm{var}}-P_i^{\mathrm{ref}}\right)^2}
{
\displaystyle\sum_{i=1}^{n}
\left(P_i^{\mathrm{ref}}\right)^2}
}
\times 100\,\% .
\end{align}
with $n$ being the total number of time steps in each loading configuration. All values were calculated with respect to the reference \texttt{tetra} formulation.

\begin{table*}[!tb]
\centering
\begin{tabular}{lrrrrrr@{\hspace{10mm}}rrrrrr}
\toprule
&
\multicolumn{6}{c}{\textbf{Peak load relative difference [\%]}\phantom{XXXX}}
&
\multicolumn{6}{c}{\textbf{NRMSE [\%]}}
\\
\textbf{simulation}
&
\texttt{C\_1}
&
\texttt{C\_1T}
&
\texttt{C\_3T}
&
\texttt{I\_1}
&
\texttt{I\_1T}
&
\texttt{I\_3T}
&
\texttt{C\_1}
&
\texttt{C\_1T}
&
\texttt{C\_3T}
&
\texttt{I\_1}
&
\texttt{I\_1T}
&
\texttt{I\_3T}
\\
\midrule

Dogbone
& 0.7 & 0.0 & 0.0 & -0.6 & 0.0 & 0.0
& 11.8 & 0.5 & 0.5 & 7.4 & 0.0 & 0.5
\\

TPB
& -0.6 & -0.5 & -0.6 & -1.5 & 0.0 & 0.2
& 6.5 & 1.7 & 1.7 & 6.2 & 0.0 & 0.3
\\

UCfixed
& \cellcolor{gray!30} -15.0 & 0.5 & 2.6 & \cellcolor{gray!30} -11.0 & 0.0 & 2.6
& \cellcolor{gray!30} 67.5 & 0.6 & 3.7 & \cellcolor{gray!30} 61.4 & 0.1 & 3.7
\\

UCfree
& \cellcolor{gray!30} -9.2 & 0.2 & 1.4 & \cellcolor{gray!30} -6.4 & 0.0 & 1.4
& \cellcolor{gray!30} 48.3 & 0.5 & 4.4 & \cellcolor{gray!30} 52.0 & 0.0 & 4.3
\\

Hardening
& - & - & - & - & - & -
& 0.2 & 0.5 & 0.6 & 0.1 & 0.5 & 0.6
\\
\bottomrule
\end{tabular}

\caption{ Comparison of peak load relative differences and normalized root mean square errors (NRMSE) with respect to the reference formulation \texttt{tetra}. Significant deviations (Peak load difference above 5\,\% and NRMSE above 20\,\%) are highlighted. \texttt{decoupledTetra} is not included as al the errors were below 0.02\,\%.
}
\label{table_errors}
\end{table*}

First, let us start with comparison of \texttt{tetra} and \texttt{decoupledTetra} models, which exhibited practically identical response in all the benchmark tests.The numerical differences remain below 0.02\,\%. These models have identical geometry and integration rule, the only difference is their internal formulation inside the solver and evaluation of volumetric strain. One can therefore verify the implementation of both of these models and conclude that supplying the volumetric strain into two-node lattice elements from auxiliary tetrahedrons in the neighborhood is a~valid technique.

With respect to the \emph{position of the central point} at the centroid or intersection, the relative difference between variants \texttt{I} and \texttt{C} is mostly negligible. Apart from the hardening test, where practically no difference occurs, the most pronounced differences in peak load are observed for the single-IP formulations,ranging from approximately 0.9\,\% in TPB to 4.0\,\% in UCfixed. For geometries with one IP per triangle, the differences remain below 0.5\,\%. For three IPs per triangle, they are generally very small as well, although the TPB test shows a~difference of approximately 0.8\,\%. The forces computed with the centroidal location are slightly lower than those computed with intersectional location. This suggests that the artificial eccentricity due to the shift of the integration point from the centroid increases the overall stiffness of the model.

The effect of the \emph{number of integration points} varies from one benchmark test to another. (i) The hardening response in confined compression (Fig.~\ref{us}) showed practically no effect of the integration rule , with NRMSE not exceeding 0.6\,\%. This suggests that bending of the interparticle contacts plays no role in this type of test. It also confirms once again that the volumetric strain evaluation in the lattice models is done correctly. Two of the other benchmark tests showed only negligible sensitivity to the integration scheme: (ii) the dogbone tension test (Fig.~\ref{dogbone}) and (iii) the three-point bending test (Fig.~\ref{tpb2}). Only minor differences appeared in the post-peak response; however, for the single-IP formulations, the accumulated difference over the complete response is reflected by NRMSE values ranging from 6.2\,\% to 11.8\,\%. The differences are attributed to rotations of the particles at the crack faces. The more integration points are included, the stiffer is the model in inter-particle bending in the crack area and therefore the stiffer is the overall response.

In contrast, the remaining benchmark tests (iv) of unconfined compression revealed pronounced differences (Fig.~\ref{uc_results}).Geometries with a~single IP reached only between 85.0\,\% and 93.6\,\% of the peak force and displayed significantly more brittle behavior. NRMSE reached up to 67.5\,\%. In comparison, integration schemes employing more than one integration point per facet produced consistent responses that were in close agreement with the reference geometries. A~small difference is also visible between the \texttt{\_1T} and \texttt{\_3T} variants in post-peak regime, with the latter exhibiting a~slightly stiffer response.This difference is also reflected in the NRMSE, which is approximately 4\,\% for the \texttt{\_3T} variants, while remaining below 0.6\,\% for the \texttt{\_1T} variants.  The results suggests that curvatures built up in unconfined compression gives rise to crucial bending couples acting at thecontacts. Single-point integration omits these couples and, therefore, completely misses the bending effects, while multi-point integration includes them. The more integration points, the better description of the traction profile in inelastic regime, but even one IP in each triangle is already quite sufficient.

These observations clearly indicate that a~single integration point is insufficient to provide the bending resistance required for the transfer of couples between aggregates, leading to an~underestimation of inter-particle bending effects. Although this limitation may be negligible in cases where inter-particle curvatures remain small, it becomes critical in loading scenarios such as unconfined compression, where bending plays a~dominant role. In such cases, the choice of an~appropriate integration rule is essential to ensure an~accurate mechanical response.

To verify this hypothesis, the entire set of virtual experiments is repeated with fully constrained particle rotations mimicking infinite bending stiffness at the contacts.. In other words, all the rotational degrees of freedom are set to zero, and only translations are left free. An~exception is made for the three-point bending, where only rotations along the longitudinal and vertical axes are restricted. Rotations around the remaining horizontal axis are left free. These rotations are essential to describe the bending character of the setup; restricting them would compromise the whole simulation.

\begin{figure}[ht!]
    \centering
    \includegraphics[trim={0 0 0 0},clip]{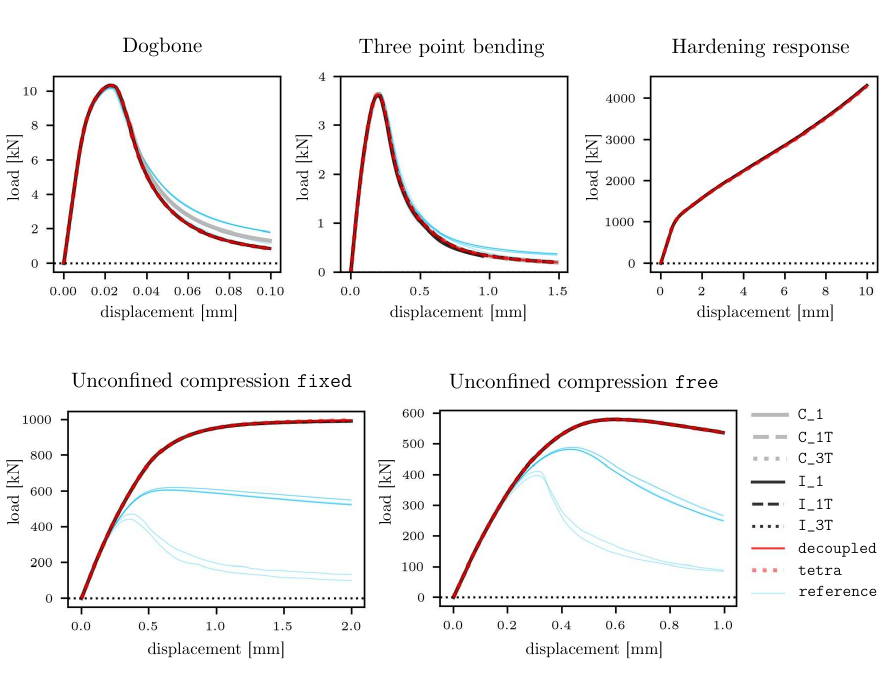}
     \caption{Load–displacement curves from simulations with restricted aggregate rotations. Comparison with the reference cases shows that suppressing rotations eliminates differences between integration schemes.}
     \label{NR}
\end{figure}

With artificially constrained rotations, the choice of integration rule should be irrelevant and should not lead to a~different sample response, because the traction is constant over the entire contact area.
Figure~\ref{NR} compares the results of these restricted-rotation simulations with the original unconstrained benchmark results, labeled as reference and plotted in light blue. Indeed, when aggregate rotations are suppressed, the choice of integration scheme no longer affects the results.One can see that such fundamental change of the model resulted in very limited consequences in simulations of three-point bending, dogbone and confined compression. This verifies that accounting for the curvature and bending at the contacts is not important in these loading scenarios. Completely opposite conclusion is drawn for unconfined compression, where curvature effects at the contacts matter significantly. 


\subsection{Computational Performance}

It is interesting to compare the execution times of different integration schemes. The same computer with processor \texttt{Intel(R) Core(TM) i9-10920X (3.50\,GHz)} and 125\,GB RAM running Ubuntu operation system was employed for all the simulations, each time only one computational thread was used. Since the implicit solver used in the simulations factorizes (LU decomposition) the initial stiffness matrix and uses it for all subsequent time steps and iteration, the majority of computational time is spent on evaluation of internal forces within the model, that is, visiting all IPs and evaluating the traction there. The convergence of all the integration schemes is similar; therefore, the execution times are expected to be closely proportional to the number of integration points in the model. A~similar comparison has been presented in \parencite{Wang2024}.

\begin{figure}[!tb]
  \centering
    \includegraphics[trim={0 0 0 0},clip,width=0.57\textwidth]{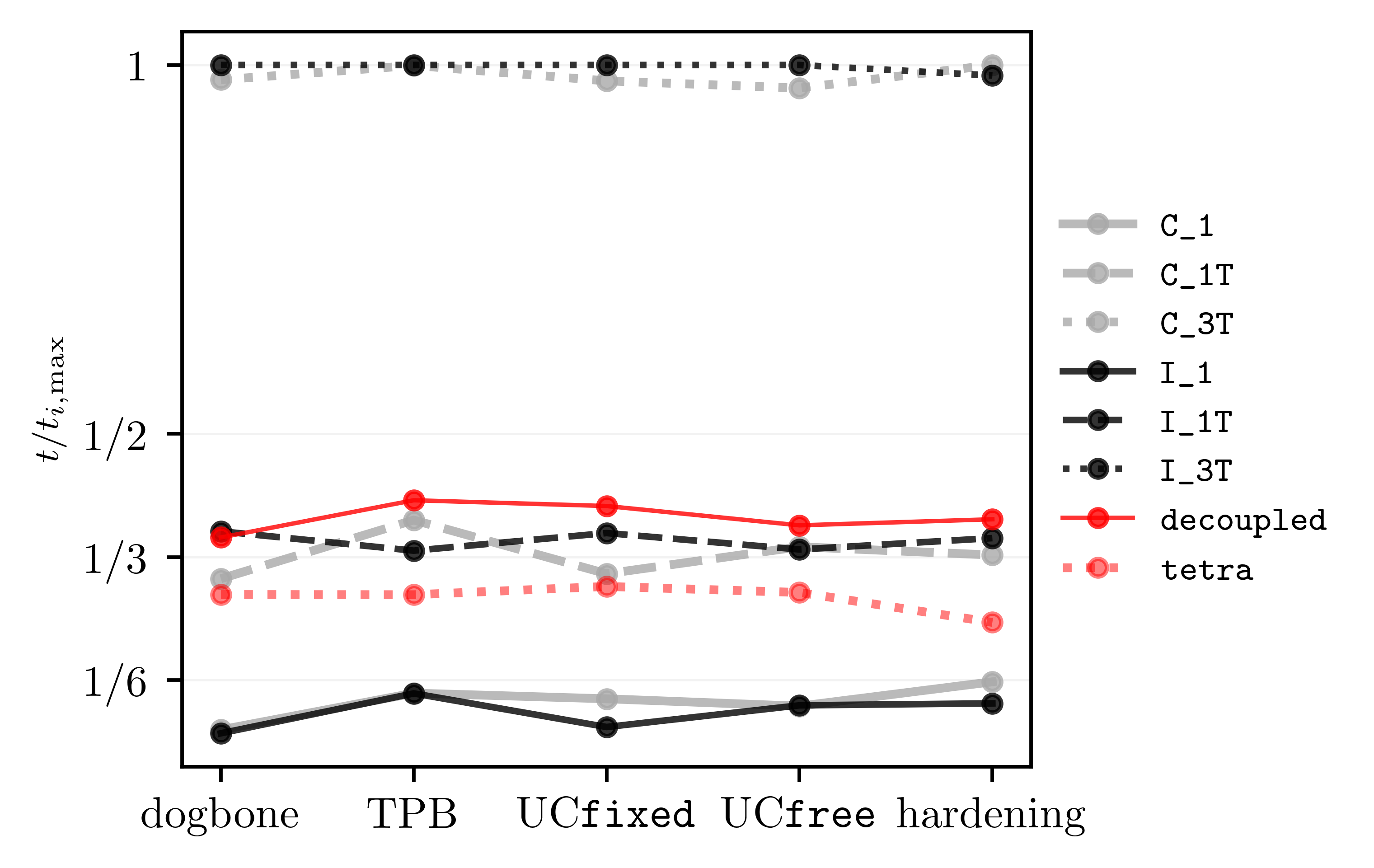}
    \caption{Comparison of computational time for different simulations and geometry configurations.}
    \label{fig:times}
\end{figure}

Two reference configurations (\texttt{tetra} and \texttt{decoupledTetra}) and six tested variants (\texttt{C\_1}, \texttt{C\_1T}, \texttt{C\_3T}, \texttt{I\_1}, \texttt{I\_1T}, \texttt{I\_3T}) are compared. The execution times, normalized with respect to the longest run of each simulation, are shown in Fig.~\ref{fig:times}. For completeness, the exact measured values and the number of integration points are provided in Tabs.~\ref{tab:performance} and \ref{tab:ips} for each simulation. The number of IPs for \texttt{C} and \texttt{I} are identical as well as for \texttt{tetra} and \texttt{decoupledTetra} are equal by the construction of the model. However,  \texttt{tetra} model has lower number of IPs compared to \texttt{C\_1, I\_1} even though the latter is created from the former. The reason lies in the boundary region where tetrahedral edge lies at the specimen boundary. In lattice elements corresponding to these edges the triangulation contains more (sometimes degenerate) triangles than the original LDPM structure. 
The data confirm the previous expectations. The integration schemes with 3 integration points in a~triangle are the most computationally expensive, while models with a~single IP in each triangle run approximately three times faster. Using only one integration point for each contact reduces the computational time even more, approximately by a~factor 3 again. One would expect a~more significant reduction here as the number of integration points has decreased approximately by factor 6-8. The difference is attributed to other operations that do not scale with the number of IPs (solution of linear system, reading and export of the data, etc.). 

Notice that the execution times of reference models \texttt{tetra} and \texttt{decoupledTetra} are similar to models with one integration point per triangle, but \texttt{decoupledTetra} seems to perform slightly worse and \texttt{tetra} is significantly better even though the number of IPs is exactly the same. This is attributed to the number of element class instances in the model. The initial number of \texttt{tetra} elements with 12 IPs is transformed to 12 times more \texttt{decoupledTetra} elements of a~single IP. When collected into lattice elements with projected facets, the number of elements is reduced, but it is still larger than the original number of \texttt{tetra} elements. For example, in the dogbone specimen simulations, there are 7\,900 \texttt{tetra} elements, 94\,800 \texttt{decoupledTetra} elements, and 10\,468 lattice \texttt{I\_1T} or \texttt{C\_1T} elements. The internal forces are evaluated in a~loop over all elements within which the IPs are visited.OAS implementation clearly favors fewer elements with more integration points. 

These findings suggest that the \texttt{\_1T} variants (the same integration rule is used in LDPM) provide the most favorable balance between computational efficiency and predictive accuracy.

\begin{table}[!tb]
\centering \begin{tabular}{lrrrrrrrr}
\toprule
&
\multicolumn{8}{c}{\textbf{Exact runtimes [h]}}
\\
&
\texttt{C\_1}
&
\texttt{I\_1}
&
\texttt{C\_1T}
&
\texttt{I\_1T}
&
\texttt{C\_3T}
&
\texttt{I\_3T}
&
\texttt{tetra}
&
\texttt{dTetra}
\\
\midrule

dogbone
& 0.34 & 0.32 & 1.04 & 1.26 & 3.36 & 3.43 & 0.97 & 1.23
\\

TPB
& 8.34 & 8.32 & 21.49 & 19.16 & 56.04 & 56.07 & 15.82 & 22.99
\\

UCfixed
& 2.67 & 1.95 & 5.88 & 6.92 & 18.53 & 18.94 & 5.55 & 7.62
\\

UCfree
& 1.08 & 1.09 & 2.85 & 2.82 & 7.96 & 8.21 & 2.34 & 3.09
\\

hardening
& 0.16 & 0.13 & 0.34 & 0.36 & 1.00 & 0.99 & 0.24 & 0.38
\\
\bottomrule
\end{tabular}

\caption{Exact runtimes for all investigated geometries and simulations.}
\label{tab:performance}
\end{table}

\begin{table}[!tb]
\centering\begin{tabular}{lrrrr}
\toprule
&
\multicolumn{4}{c}{\textbf{Number of IPs}}
\\
&
\texttt{C\_1}, \texttt{I\_1}
&
\texttt{C\_1T}, \texttt{I\_1T}
&
\texttt{C\_3T}, \texttt{I\_3T}
&
\texttt{tetra}, \texttt{dTetra}
\\
\midrule

dogbone
& 10\,468 & 99\,977 & 299\,931 & 94\,800
\\

TPB (discrete part)
& 17\,534 & 167\,604 & 502\,812 & 158\,964
\\

TPB (discrete + FEM)
& 92\,486 & 242\,556 & 577\,764 & 233\,916 
\\

UCfixed
&16\,234 & 158\,034 & 474\,102 & 151\,644
\\

UCfree
& 16\,234 & 158\,034 & 474\,102 & 151\,644
\\

hardening
& 11\,628 & 113\,136 & 339\,408 & 108\,576
\\
\bottomrule
\end{tabular}
\caption{Number of integration points for all investigated geometries and simulations. Values for TPB are reported for the discrete part only and for both the discrete and FEM parts together.}
\label{tab:ips}
\end{table}

\section{Conclusion}

Various integration schemes in discrete models for simulating mechanical behavior of cohesive heterogeneous materials are discussed in the article. The fundamental difference between these schemes is manifested by the rotational stiffness of the contacts. The resistance to curvature increases with the number of integration points. The Lattice Discrete Particle Model (LDPM) with tetrahedral elements serves as a~reference framework. To investigate the integration rules, LDPM geometry is reformulated into lattice-type two-node elements with projected polygonal area. Three different integration rules were considered: (i) single integration point for the whole 
 polygonal contact face located at its central point, (ii) one integration point at the centroid of each triangle created from the central point and a~line segment at the contact boundary, and (iii) three integration points per each triangle. On top of that, two locations of the central point were tested: either at the centroid of the contact area or at the point where the branch vector intersects the contact plane. The former variant is more representative and is often used in lattice-particle models, while the latter one emerges from the LDPM geometry transformation. In addition to the mechanical response of the discrete system, the effect of integration scheme was studied also from the point of view of computational times.

Analysis of benchmark fracture simulations, including dogbone tension test, three-point bending, unconfined compression, and hardening response, showed that the tensile, bending, and hardening behavior was largely unaffected by the choice of the integration scheme. In contrast, unconfined compression proved to be highly sensitive: single-point integration underestimated the peak load and produced a~more brittle response, whereas multi-point schemes delivered consistent results comparable with the reference LDPM geometry. Additional artificially modified simulations with fully constrained rotations confirmed that these differences originate from the role of emerging bending resistance in the transmission of  couples between particles. Increasing the number of particles in the system would not improve the unconfined compressive response, as the transfer of bending moments at the contacts would still be absent. Instead, rotational stiffness must be introduced by accounting for a~non-constant traction profile through multiple integration points. In principle, one could compensate for the missing bending contribution in a~single-point interaction by assigning artificially high strengths and fracture energies to the contacts. Such a~workaround would, however, fail once the model is applied to predominantly tensile loading scenarios. This conjecture should be verified in future work by comparing model predictions with an~experimental campaign covering a~range of loading scenarios using the same material. 

The computational assessment showed that execution time increases with the number of integration points. A~specific implementation of the steady-state implicit solver ensured that majority of the computational time is spent on evaluation of internal forces, which is directly proportional to the number of integration points. Therefore, models with three integration points per triangle needed three times longer execution times compared to models with one integration point per triangle. This huge run-time difference of multi-point integration schemes results in almost negligible differences in mechanical responses. Single integration point schemes were fastest, but provided incorrect response to unconfined compression.

Finally, concrete behavior under confined compression is determined by volumetric strain. LDPM, formulated as tetrahedral elements connecting four particles, is capable of computing the volumetric strain from the change of volume of the tetrahedron. In a~lattice system that only features two-node contacts, the volumetric strain cannot be evaluated at the level of elements. It has to be supplied from the surrounding volume, in the present case from auxiliary tetrahedrons constructed in the domain specifically to evaluate the volumetric strains. 

In summary, while the tensile and bending responses are governed primarily by normal and shear interactions, the compression-dominated behavior depends strongly on the ability of the model to transmit bending between aggregates. Integration schemes with a~single point per contact underestimate this mechanism and lead to overly soft responses, whereas multi-point schemes provide consistent predictions. The location of the central point (intersection or centroid) is practically irrelevant in all benchmark tests. The volumetric strain needed in the constitutive function describing confined compression has to be supplied from the neighborhood of the two-node contacts, for example, by tetrahedralization of the domain and averaging the volumetric strains from adjacent tetrahedrons. 

\section*{Acknowledgement} This research work was funded by the National Science Centre, Poland under the OPUS call in the Weave programme project no. 2023/51/I/ST11/00069 and by the Czech Science Foundation project no. 25-14337L for the bilateral project, titled ``Multi-fidelity modelling of shear concrete strength based on experiments, simulation and probabilistic approaches (MMOSS)''. The simulations were performed  on computational servers acquired within project INODIN (Innovative methods of materials diagnostics and monitoring of engineering infrastructure to increase its durability and service time – CZ.02.01.01/00/23\_020/0008487) co-funded by European Union.

\appendix
\section*{Appendix A: Constitutive model and material parameters}
\label{app:constitutive}

For completeness, the full constitutive formulation of the Lattice Discrete Particle Model (LDPM) \parencite{CusMen-11} updated by Ref.~\parencite{CecSal-17} is provided here. 

The sign of the normal strain $e_N$ determines whether a~facet is in tension or compression. In the case of tension, fracturing behavior is defined in the following manner. Effective strain $\eeff$ and effective traction $\teff$ are defined
\begin{align}
\eeff &= \sqrt{ e_N^{2} + \alpha(e_M^{2} + e_L^{2}) } &
\teff &= \sqrt{ t_N^{2} + (t_M^{2} + t_L^{2}) / \alpha }
\end{align}
The effective traction has to satisfy the inequality $0 \leq \teff \leq \sigma_{bt}(\varepsilon, \omega)$ where $\sigma_{bt}(\varepsilon, \omega)$ is a~limiting boundary depending on the effective strain $\eeff$ and the parameter $\omega$ which represents the interaction between normal and shear loading.
\begin{align}
\tan \omega &= \frac{e_N}{\sqrt{\alpha\left(e_M^{2} + e_L^{2}\right)}} &
\sigma_{bt}(e_{\max}, \omega) &= \sigma_{0}(\omega) \exp \left[ -H_{0}(\omega) 
\frac{ \mathrm{max}\{0, e_{\max} - \sigma_{0}(\omega)/E_0\} }{ \sigma_{0}(\omega) } \right]
\end{align}
$e_{\max}$ is the maximum effective strain obtained in history. The function $\sigma_{0}(\omega)$ is the strength limit for effective traction and $H_{0}(\omega)$ governs the softening
\begin{align}
\sigma_{0}(\omega) &= \sigma_{t} 
\frac{ -\sin(\omega) + \sqrt{ \sin^{2}(\omega) + 4 \alpha \cos^{2}(\omega) / r_{st}^{2} } }
{ 2 \alpha \cos^{2}(\omega) / r_{st}^{2} } 
& 
H_{0}(\omega) &= H_{t} \left( \frac{2 \omega}{\pi} \right)^{n_{t}}
\end{align}
where $r_{st} = \sigma_{s}/\sigma_{t}$ represents cohesion as the ratio of shear strength $\sigma_{s}$ to tensile strength $\sigma_{t}$, $n_{t}$ is a~softening exponent calibrated from experiments, the softening modulus is $H_{t} = 2E_{0}/(l_{t}/l-1)$, and $l_{t} = 2E_{0}G_{t}/\sigma_{t}^{2}$, $G_{t}$ is the mesoscale fracture energy, and $l$ is the distance between centers of particles in contact.

In compression the constitutive relations aim to reproduce the experimentally observed behavior of pore collapse and material densification.  A~boundary $\sigma_{bc}(\varepsilon_{D}, \varepsilon_{V})$ is employed as a~function of deviatoric strain $\varepsilon_{D}$ and volumetric strain $\varepsilon_{V}$, expressed by their ratio $r_{DV}$. An~updated version from Ref.~\parencite{CecSal-17} is used to define the compressive behavior
\begin{align}
    r_{DV} =
\begin{cases}
-\frac{|\varepsilon_{D}|}{\varepsilon_{V} - \varepsilon_{V0}}, & \text{for } \varepsilon_{V} \leq 0, \\
\frac{|\varepsilon_{D}|}{\varepsilon_{V0}}, & \text{for } \varepsilon_{V} > 0 .
\end{cases}
\end{align}
where $\varepsilon_{V0} = \kappa_{c3} \cdot \sigma_{0}(\omega)/E_{0}$. Here $\kappa_{c3} = 0.1$  based on experimental data. The boundary is defined as follows:
\begin{align}
\sigma_{bc}(\varepsilon_{D}, \varepsilon_{V}) =
\begin{cases}
\sigma_{c0} & \text{for } -\varepsilon_{DV} \leq 0 \\
\sigma_{c0} + \langle -\varepsilon_{DV} - \varepsilon_{c0} \rangle H_{c}(r_{DV}) & \text{for } 0 \leq -\varepsilon_{DV} \leq \varepsilon_{c1}, \\
\sigma_{c1}(r_{DV}) \exp \left[ \frac{(-\varepsilon_{DV} - \varepsilon_{c1}) H_{c}(r_{DV})}{\sigma_{c1}(r_{DV})} \right],& \text{otherwise}.
\end{cases}
\end{align}
Here the compaction strain is $\varepsilon_{DV} = \varepsilon_{V} + \beta \varepsilon_{D}$, $\varepsilon_{c0} = \sigma_{c0}/E_{0}$ is the compaction strain when pores start to collapse at yielding compressive traction, $\sigma_{c1}(r_{DV})$ is the compressive strength at the onset of rehardening, $H_{c}(r_{DV})$ is the hardening modulus, and $\varepsilon_{c1} = \kappa_{c0} \varepsilon_{c0}$ is the compaction strain at which rehardening begins, with $\sigma_{c1}(r_{DV}) = \sigma_{c0} + (\varepsilon_{c1} - \varepsilon_{c0}) H_{c}(r_{DV})$.  The hardening modulus $H_{c}$ is given by:
\begin{align}
H_{c} = \frac{H_{c0} - H_{c1}}{1 + \kappa_{c2} \langle r_{DV} - \kappa_{c1} \rangle} + H_{c1},
\end{align}
where $\beta$, $H_{c0}$, $H_{c1}$, $\kappa_{c1}$, and $\kappa_{c2}$ are material parameters defined from experimental data.

Frictional behavior under compression is governed by the following formula 
\begin{align}
\sigma_{bs} = \sigma_{s} + (\mu_{0} - \mu_{\infty}) \sigma_{N0} - \mu_{\infty}t_N 
- (\mu_{0} - \mu_{\infty}) \sigma_{N0} \exp\!\left(t_N/\sigma_{N0}\right),
\end{align}
where $\sigma_{s}$ is the cohesion and $\sigma_{N0}$ is the traction at which the transition between the initial friction coefficient $\mu_{0}$ and the final friction coefficient $\mu_{\infty}$.

A complete account of the constitutive law can be found in \parencite{CusMen-11,CecSal-17}.All material parameters used in the present study were taken from \parencite{CusMen-11}; no parameters were measured or calibrated specifically for this work. In \parencite{CusMen-11}, the parameters were calibrated against a~three-point bending test, where a~detailed description of the calibration procedure can be found. The parameter values adopted in the present study are listed in Table \ref{table_params}.

\begin{table}[htbp]
\centering
\begin{tabular}{llll}
\toprule
\textbf{Parameter} & \textbf{Symbol} & \textbf{Value} & \textbf{Unit} \\
\midrule
Normal Modulus & $E_{0}$ & 60273 & MPa \\
Alpha & $\alpha$ & 0.25 & -- \\
Tensile Strength & $\sigma_{t}$ & 3.44 & MPa \\
Tensile Characteristic Length & $l_{t}$ & 500 & mm \\
Fracture Energy & $G_{t}$ or $(l_{t}\sigma_{t}^{2}/(2E_{0}))$ & 0.0491 & N/mm \\
Shear Strength Ratio & $r_{t}$ or $(\sigma_{s}/\sigma_{t})$ & 2.6 & -- \\
Softening Exponent & $n_{t}$ & 0.4 & -- \\
Compressive Yielding Strength & $\sigma_{c0}$ & 150 & MPa \\
Initial Hardening Modulus Ratio & $H_{c0}/E_{0}$ & 0.40 & -- \\
Transitional Strain Ratio & $\kappa_{c0}$ & 4 & -- \\
Deviatoric Strain Threshold Ratio & $\kappa_{c1}$ & 1 & -- \\
Deviatoric Damage Parameter & $\kappa_{c2}$ & 5 & -- \\
Initial Friction & $\mu_{0}$ & 0.4 & -- \\
Asymptotic Friction & $\mu_{\infty}$ & 0 & -- \\
Transitional traction & $\sigma_{N0}$ & 600 & MPa \\
Densification Ratio & $E_{d}/E_{0}$ & 1 & -- \\
Volumetric Deviatoric Coupling & $\beta$ & 0 & -- \\
Final Hardening Modulus Ratio & $H_{c1}/E_{0}$ & 0.1 & -- \\\bottomrule
\end{tabular}
\caption{Material parameters used in the experiments.}
\label{table_params}
\end{table}

\printbibliography

\end{document}